\documentstyle[preprint,aps,epsf]{revtex}

\newcommand{\sla}{\!\!\!\!/ \,}

\begin{document}

\newcommand{\bea}{\begin{eqnarray}}
\newcommand{\eea}{\end{eqnarray}}

\title {Evaluating Real Time Finite Temperature Feynman Amplitudes}

\author
{M.E. Carrington${}^{a,b}$, Hou 
Defu${}^{a,b,c}$, A. Hachkowski${}^{a,d}$, D. Pickering${}^{e}$ and J.C. Sowiak${}^{a,e}$}

\address{
${}^a$ Department of Physics, Brandon University, 
Brandon, Manitoba, R7A 6A9 
Canada\\
${}^b$  Winnipeg Institute for Theoretical Physics, Winnipeg, Manitoba\\
${}^c$  Institute of Particle Physics, Huazhing Normal University, 430070 Wuhan, China\\
${}^d$ Current address: Faculty of Engineering, 
University of Manitoba, Winnipeg, Manitoba, R3T 5V6 Canada \\
${}^e$ Department of Mathematics, Brandon University, Brandon, Manitoba, R7A 6A9 Canada\\
}

\date{\today}
\maketitle

\begin{abstract}
We construct a program to calculate Feynman amplitudes
at finite temperature in the real time Keldysh formalism using the symbolic manipulation 
program {\it  Mathematica}. As an example, the usefulness of this program is demonstrated by proving the finite temperature Ward identity for
 QED in a second order effective theory.
\end{abstract}

\pacs{PACS numbers: 11.10Wx, 11.15Tk, 11.55Fv}
\date{\today}

\vspace{3ex}

\section{Introduction}

There are two different methods for calculating Feynman amplitudes at finite temperature.  
Each of these methods presents technical difficulties\cite{kap,leb,rep-145}.  The imaginary time formalism 
involves the calculation of green functions with imaginary time arguments.  At the end 
of the calculation, one must perform an analytic continuation to obtain real time green 
functions.  For higher n-point functions, these analytic continuations become increasingly 
difficult.  In spite of this difficulty, the imaginary time formalism has been traditionally 
the most popular.

The difficulty associated with the real time formalism is the doubling of degrees of freedom. 
The real time integration contour involves two branches, one running from minus infinity to 
positive infinity just above the real axis, and one running back from positive infinity to 
negative infinity just below the real axis\cite{Chou,PeterH}.  All fields can take values on either branch of 
the contour and thus there is a doubling of the number of degrees of freedom.  For example, 
the two point function becomes a two by two matrix.  The four components of this matrix 
 represent the four possible contractions of two field operators each of which can take values 
on either of the two branches of the real time contour. The physical two point functions 
(for example, the retarded and advanced two-point functions) can be extracted by taking 
apropriate combinations of the four components of this matrix.  The procedure is 
similar for higher n-point functions.  It is straightforward to show that this doubling 
of degrees of freedom is necessary to obtain finite green functions. 

The problem with the real time formalism is that the extra degrees of freedom become 
increasingly cumbersome to handle for higher n-point functions.  Each line within a given Feynman diagram contains a Keldysh index at each end which takes values $\{1,2\}$, corresponding to the two branches of the closed time path contour.  Indices that do not correspond to external legs are called internal indices, and must be summed over.  The indices that correspond to external legs are called external indices.  As a consequence of these indices, an n-point function has $2^n$ components.  These components obey one 
constraint equation, which reduces the number of independent components to $2^n-1$.  In 
equilibrium, the KMS conditions impose additional constraints, reducing the number of 
independent components to $2^{n-1} -1$.

In summary then, the imaginary time formalism has the advantage that the initial calculation 
is easier, but the disadvantage that analytic continuations must be done, which makes it 
difficult to extract the physical green functions.  In contrast, in the real time formalism, 
the initial calculation is more difficult, but there is a simple and natural procedure for 
extracting the physical green functions.  In addition, the real time formalism can be generalized to non-equilibrium situations.  In this paper we describe a $Mathematica$ calculation that does sums over internal indices, and takes physical combinations of external indices.  Using this program substantially reduces the technical difficulties associated with the real time formalism, and makes it possible for us to exploit its advantages.  

This paper is organized as follows:  In the section II we discuss the definitions of real time finite temperature green functions within the Keldysh formalism, and define our notation.  In the section III, we describe the $Mathematica$ program that we have written to do summations over internal Keldysh indices, and take physical combinations of external indices.  In the  section III, we demonstrate the usefulness of this program by using it to prove the Ward identity for QED in a second order hard thermal loop (HTL) effective theory.  In the last section, we give some conclusions.  

The program and the manuscript of this paper are available on the Los Alamos data base (hep-ph/9908438) and at http://theory.uwinnipeg.ca/users/hdf

\section{Real Time Green Functions}

\subsection{The two-point function}

We first consider the
propagator. In real time, the propagator has $2^2=4$ components, since
each of the two fields can take values on either branch of the
contour. Thus, the propagator can be written as a $2 \times 2 $ matrix
of the form
 \begin{equation}
 \label{2}
   D = \left(  \matrix {D_{11} & D_{12} \cr
                        D_{21} & D_{22} \cr} \right) \, ,
 \end{equation}
where $D_{11}$ is the propagator for fields moving along $C_1$,
$D_{12}$ is the propagator for fields moving from $C_1$ to $C_2$, etc.
The four components are given by
 \begin{eqnarray}
 \label{eq: compD}
   D_{11}(x-y) &=& -i\langle T(\phi(x) \phi(y))\rangle \, , \nonumber\\
   D_{12}(x-y) &=& -i\langle \phi(y) \phi(x) \rangle \, , \nonumber\\
   D_{21}(x-y) &=& -i\langle \phi(x) \phi(y)\rangle \, , \nonumber\\
   D_{22}(x-y) &=& -i\langle\tilde{T}(\phi(x)\phi(y))\rangle \, ,
 \end{eqnarray}
where $T$ is the usual time ordering operator, and $\tilde{T}$ is the
anti-chronological time ordering operator. These four components
satisfy,
 \begin{equation}
 \label{3}
   D_{11} - D_{12} - D_{21} + D_{22} = 0
 \end{equation}
as a consequence of the identity $\theta(x) + \theta(-x) =1$.

It is more useful to write the propagator in terms of the three functions
 \begin{eqnarray} \label{3a}
   D_R &=& D_{11} - D_{12} \, , \nonumber\\
   D_A &=& D_{11} - D_{21} \, , \nonumber\\
   D_F &=& D_{11} + D_{22} \, .
 \end{eqnarray}
$D_R$ and $D_A$ are the usual retarded and advanced propagators,
satisfying
 \begin{equation}
 \label{4}
   D_R(x-y)-D_A(x-y) = -i\langle [\phi(x),\phi(y)] \rangle\, ,
 \end{equation}
and $D_F$ is the symmetric combination
 \begin{equation}
 \label{4a}
   D_F(x-y) = -i\langle \{\phi(x),\phi(y)\} \rangle\, .
 \end{equation}
In momentum space these three propagators satisfy the KMS condition, 
\begin{equation}
D_F(P) = (1+2n(p_0))(D_R(P) - D_A(P)),
\label{eq: 4b}
\end{equation}
where
 \begin{equation}
\label{DRA}
   D_{R,A}(P) = \frac{1}{(p_0\pm i\epsilon)^2 - {\vec p}^{\,2}-m^2}
 \end{equation}
and $n(p_0)$ is the Bose-Einstein distribution function,
\bea
n(p_0) = \frac{1}{e^{\beta p_0}-1}
\label{BE}
\eea

Equations~(\ref{3}),~(\ref{3a}) are inverted by
 \begin{eqnarray}
 \label{7}
   D_{11} &=& \frac{1}{2} (D_F + D_A + D_R) \, , \nonumber\\
   D_{12} &=& \frac{1}{2} (D_F  +D_A - D_R) \, , \nonumber\\
   D_{21} &=& \frac{1}{2} (D_F  -D_A + D_R) \, , \nonumber\\
   D_{22} &=& \frac{1}{2} (D_F  -D_A - D_R) \, .
 \end{eqnarray}
These equations can be written in a more convenient notation as \cite{Chou}
 \begin{equation}
 \label{decompD}
   2\,D = D_R {1\choose 1}{1\choose -1}
        + D_A {1\choose -1}{1\choose 1}
        + D_F {1\choose 1}{1\choose 1}
 \end{equation}
where the outer product of the column vectors is to be taken.
\par
Similar relations can be obtained for the 1PI two-point function, or the 
polarization tensor, which is obtained by amputating the external legs from
the propagator. The Dyson equation gives
 \begin{equation}
 \label{8}
   iD(p) = iD_0(p) + iD_0(p) \, \bigl(-i\Pi(p)\bigr)\, iD(p) \, .
 \end{equation}
The analogues of~(\ref{3}) and~(\ref{3a}) are
 \begin{eqnarray}
   \Pi_R &=& \Pi_{11} + \Pi_{12} \, , \nonumber\\
   \Pi_A &=& \Pi_{11} + \Pi_{21} \, , \nonumber\\
   \Pi_F &=& \Pi_{11} + \Pi_{22} \, ,
 \label{physpi}
 \end{eqnarray}
and
 \begin{equation}
   \Pi_{11} + \Pi_{12} + \Pi_{21} + \Pi_{22} = 0 \, .
 \label{eq: circpi}
 \end{equation}
The analogues of~(\ref{decompD}) and~(\ref{eq: 4b}) are
 \begin{eqnarray}
 \label{eq: Pidecomp1}
   2\, \Pi (p) &=& \Pi_R(p) {1\choose -1} {1\choose 1}
             + \Pi_A(p) {1\choose 1} {1\choose -1}
             + \Pi_F(p) {1\choose -1} {1\choose -1}\, ,
 \\
 \label{eq: KMSpi}
    \Pi_F(p) &=& \Bigl( 1+2n(p_0) \Bigr)\,
                 \Bigl( \Pi_R(p) - \Pi_A(p) \Bigr) \, .
  \end{eqnarray}

\subsection{Three-Point Function}
\label{sec5}

In the real time formalism, the three-point function has $2^3 = 8$
components. We denote the connected functions
by $\Gamma^C_{abc}$ where $\{a,b,c = 1,2\}$.
In analogy to~(\ref{eq: compD}), they are given by the following
expressions \cite{Chou}:
 \begin{eqnarray}
   \Gamma^C_{111}(x,y,z)
    &=& \langle T(\phi(x)\phi(y)\phi(z))\rangle \, , \nonumber\\
   \Gamma^C_{112}(x,y,z)
    &=& \langle\phi(z)\,T(\phi(x)\phi(y))\rangle \, , \nonumber\\
   \Gamma^C_{121}(x,y,z)
    &=& \langle\phi(y)\,T(\phi(x)\phi(z))\rangle \, , \nonumber\\
   \Gamma^C_{211}(x,y,z)
    &=& \langle\phi(x)\,T(\phi(y)\phi(z))\rangle \, , \nonumber\\
   \Gamma^C_{122}(x,y,z)
    &=& \langle\tilde{T}(\phi(y)\phi(z))\,\phi(x)\rangle\, , \nonumber\\
   \Gamma^C_{212}(x,y,z)
    &=& \langle\tilde{T}(\phi(x)\phi(z))\,\phi(y)\rangle\, , \nonumber\\
   \Gamma^C_{221}(x,y,z)
    &=& \langle\tilde{T}(\phi(x)\phi(y))\,\phi(z)\rangle\, , \nonumber\\
   \Gamma^C_{222}(x,y,z)
    &=& \langle\tilde{T}(\phi(x)\phi(y)\phi(z))\rangle\, .
 \label{eq: compVer}
 \end{eqnarray}
It is immediately obvious that one of these components is dependent on the others 
because of the identity
 \begin{equation}
   \sum_{a=1}^2\sum_{b=1}^2\sum_{c=1}^2
   (-1)^{a+b+c-3} \Gamma^C_{abc} =0
 \label{eq: circVer}
 \end{equation}
which follows in the same way as~(\ref{3}) from $\theta(x) +
\theta(-x) = 1$. The seven combinations that we use are
defined as \cite{Chou}
 \begin{eqnarray}
   \Gamma^C_{R} &=& \Gamma^C_{111} - \Gamma^C_{112}
                - \Gamma^C_{211} + \Gamma^C_{212},
 \nonumber\\
   \Gamma^C_{Ri} &=& \Gamma^C_{111} - \Gamma^C_{112}
                   - \Gamma^C_{121} + \Gamma^C_{122},
 \nonumber\\
   \Gamma^C_{Ro} &=& \Gamma^C_{111} - \Gamma^C_{121}
                   - \Gamma^C_{211} + \Gamma^C_{221},
 \nonumber\\
   \Gamma^C_{F} &=& \Gamma^C_{111} - \Gamma^C_{121}
                + \Gamma^C_{212} - \Gamma^C_{222},
 \nonumber\\
   \Gamma^C_{Fi} &=& \Gamma^C_{111} + \Gamma^C_{122}
                   - \Gamma^C_{211} - \Gamma^C_{222},
 \nonumber\\
   \Gamma^C_{Fo} &=& \Gamma^C_{111} - \Gamma^C_{112}
                   + \Gamma^C_{221} - \Gamma^C_{222},
 \nonumber\\
   \Gamma^C_{E} &=& \Gamma^C_{111} + \Gamma^C_{122}
                + \Gamma^C_{212} + \Gamma^C_{221}.
 \label{eq: physVer}
 \end{eqnarray}
In coordinate space we always label the first leg of the three-point
function by $x$ and call it the ``incoming leg $(i)$", the third leg we
label by $z$ and call it the ``outgoing leg $(o)$", and the second
(middle) leg we label by $y$. Inserting the definitions~(\ref{eq:
compVer}) into~(\ref{eq: physVer}) one finds
 \begin{eqnarray}
   \Gamma^C_{R} &=&
     \theta_{23} \theta_{31} \langle [[\phi_2,\phi_3], \phi_1] \rangle
   + \theta_{21} \theta_{13} \langle [[\phi_2,\phi_1], \phi_3] \rangle \, ,
 \nonumber\\
   \Gamma^C_{Ri} &=&
     \theta_{12} \theta_{23} \langle [[\phi_1,\phi_2], \phi_3] \rangle
   + \theta_{13} \theta_{32} \langle [[\phi_1,\phi_3], \phi_2] \rangle \, ,
 \nonumber\\
   \Gamma^C_{Ro} &=&
     \theta_{32} \theta_{21} \langle [[\phi_3,\phi_2], \phi_1] \rangle
   + \theta_{31} \theta_{12} \langle [[\phi_3,\phi_1], \phi_2] \rangle \, ,
 \nonumber\\
   \Gamma^C_{F} &=&
     \theta_{12} \theta_{23} \langle \{ [\phi_1,\phi_2], \phi_3\} \rangle
   + \theta_{32} \theta_{21} \langle \{ [\phi_3,\phi_2], \phi_1\} \rangle
   + \theta_{12} \theta_{32} \langle [ \{ \phi_3,\phi_1\}, \phi_2] \rangle
   \, ,
 \nonumber\\
   \Gamma^C_{Fi} &=&
     \theta_{21} \theta_{13} \langle \{ [\phi_2,\phi_1], \phi_3\} \rangle
   + \theta_{31} \theta_{12} \langle \{ [\phi_3,\phi_1], \phi_2\} \rangle
   + \theta_{21} \theta_{31} \langle [ \{ \phi_2,\phi_3\}, \phi_1] \rangle
   \, ,
 \nonumber\\
   \Gamma^C_{Fo} &=&
     \theta_{23} \theta_{31} \langle \{ [\phi_2,\phi_3], \phi_1\} \rangle
   + \theta_{13} \theta_{32} \langle \{ [\phi_1,\phi_3], \phi_2\} \rangle
   + \theta_{13} \theta_{23} \langle [ \{ \phi_1,\phi_2\}, \phi_3] \rangle
   \, ,
 \nonumber\\
   \Gamma^C_E &=&
     \theta_{13} \theta_{23} \langle \{ \{\phi_1,\phi_2\}, \phi_3\} \rangle
   + \theta_{21} \theta_{31} \langle \{ \{\phi_2,\phi_3\}, \phi_1\} \rangle
   + \theta_{12} \theta_{32} \langle \{ \{\phi_1,\phi_3\}, \phi_2\} \rangle
   \, ,
 \label{commutators}
 \end{eqnarray}
where we have used the obvious shorthands $\phi_1 \equiv \phi(x)$,
$\phi_2 \equiv \phi(y)$, $\phi_3 \equiv \phi(z)$, and $\theta_{12}
\equiv \theta(x_0-y_0)$, etc. The first three are the retarded three-point
functions; $\Gamma^C_{Ri}$ is retarded with respect to $x_0$,
$\Gamma^C_{Ro}$ is retarded with respect to $z_0$, and $\Gamma^C_R$ is
retarded with respect to $y_0$.

The 1PI vertex functions are obtained from the connected functions by truncating 
external legs.  We will denote 1PI vertex functions by ${\Gamma}$.  
 We can write ${\Gamma}$ as a tensor of the form
\begin{equation}
{\Gamma} = { x \choose y} { u\choose v} {w \choose z}
\end{equation} where the outer product of the column vectors is to be taken. 
For the 1PI functions the analogues of~(\ref{eq: circVer}) and~(\ref{eq: physVer}) are, 
 \begin{equation}
   \sum_{a=1}^2\sum_{b=1}^2\sum_{c=1}^2
    {\Gamma}_{abc} = x+y+u+v+w+z  =0,
 \label{eq: circVer2}
 \end{equation}
and,
 \begin{eqnarray}
   {\Gamma}_{R} &=& {\Gamma}_{111} + {\Gamma}_{112}
                + {\Gamma}_{211} + {\Gamma}_{212} \, = \frac{1}{2}(x+y)(u-v)(w+z),
 \nonumber\\
  {\Gamma}_{Ri} &=& {\Gamma}_{111} + {\Gamma}_{112}
                   + {\Gamma}_{121} + {\Gamma}_{122} \, = \frac{1}{2}(x-y)(u+v)(w+z),
 \nonumber\\
   {\Gamma}_{Ro} &=& {\Gamma}_{111} + {\Gamma}_{121}
                   + {\Gamma}_{211} + {\Gamma}_{221} \,=\frac{1}{2}(x+y)(u+v)(w-z) ,
 \nonumber\\
   {\Gamma}_{F} &=& {\Gamma}_{111} + {\Gamma}_{121}
                + {\Gamma}_{212} + {\Gamma}_{222} \, = \frac{1}{2}(x-y)(u+v)(w-z) ,
 \nonumber\\
   {\Gamma}_{Fi} &=& {\Gamma}_{111} + {\Gamma}_{122}
                   + {\Gamma}_{211} + {\Gamma}_{222} \, = \frac{1}{2}(x+y)(u-v)((w-z) ,
 \nonumber\\
   {\Gamma}_{Fo} &=& {\Gamma}_{111} + {\Gamma}_{112}
                   + {\Gamma}_{221} + {\Gamma}_{222} \, = \frac{1}{2}(x-y)(u-v)(w+z) ,
 \nonumber\\
   {\Gamma}_{E} &=& {\Gamma}_{111} + {\Gamma}_{122}
                + {\Gamma}_{212} + {\Gamma}_{221} \, =\frac{1}{2}(x-y)(u-v)(w-z).
 \label{eq: physVer2}
 \end{eqnarray}
The 1PI vertex functions 
${\Gamma}(P_1,P_2,P_3)$ are related to the connected vertex functions $\Gamma^C(P_1,P_2,P_3)$ as follows: 
\begin{eqnarray}
\Gamma^C_{R} &=& i^3 a_1 r_2 a_3 {\Gamma}_{R}\nonumber \\
\Gamma^C_{Ri} &=& i^3 r_1 a_2 a_3 {\Gamma}_{Ri} \nonumber \\
\Gamma^C_{Ro} &=& i^3 a_1 a_2 r_3 {\Gamma}_{Ro} \nonumber \\
\Gamma^C_{F} &=& i^3[ r_1 a_2 f_3 {\Gamma}_{Ri}
+  f_1 a_2 r_3 {\Gamma}_{Ro} +  r_1 a_2 r_3 {\Gamma}_{F}] \nonumber \\
\Gamma^C_{Fi} &=& i^3 [ a_1 r_2 f_3 {\Gamma}_{R} 
+  a_1 f_2 r_3 {\Gamma}_{Ro} 
+ a_1 r_2 r_3 {\Gamma}_{Fi}] \nonumber \\
\Gamma^C_{Fo} &=& i^3 [ r_1 f_2 a_3 {\Gamma}_{Ri} 
+  f_1 r_2 a_3 {\Gamma}_{R}
+  r_1 r_2 a_3 {\Gamma}_{Fo}] \nonumber \\
{\Gamma}^C_{E} &=& i^3 [ 
f_1 r_2 f_3 {\Gamma}_{R} +  r_1 f_2 f_3  {\Gamma}_{Ri} +  
f_1 f_2 r_3 {\Gamma}_{Ro} \nonumber \\ 
&& \hspace*{2.2cm} +   
r_1 f_2 r_3 {\Gamma}_{F} +  
f_1 r_2 r_3 {\Gamma}_{Fi} +  
r_1 r_2 f_3 {\Gamma}_{Fo} 
+ r_1 r_2 r_3 {\Gamma}_{E} ] \nonumber 
\end{eqnarray}
where we have used the notation $D_R(P_1) = r_1$, $D_F(P_2) = f_2$, etc.

For calculational purposes, we want to obtain a decomposition of
the 1PI three-point function in terms of the seven functions~(\ref{eq: physVer2}), 
in analogy to~(\ref{decompD}) for the two-point
function. Inverting~(\ref{eq: physVer2}) we obtain
 \begin{eqnarray}
   &&4\,{\Gamma} = {\Gamma}_R {1 \choose 1} {1\choose -1} {1\choose 1}
               + {\Gamma}_{Ri} {1 \choose -1}{1\choose 1} {1 \choose 1} + 
                {\Gamma}_{Ro} {1 \choose 1} {1\choose 1} {1 \choose -1}
\label{decompG} \\             &&+ {\Gamma}_F {1\choose -1} {1\choose 1}{1\choose -1}
+ {\Gamma}_{Fi} {1\choose 1}{1\choose -1}{1\choose -1}
             + {\Gamma}_{Fo} {1\choose -1}{1\choose -1}{1\choose 1}
+ {\Gamma}_E {1\choose -1}{1\choose -1}{1\choose -1}
 \nonumber
 \end{eqnarray}

\subsection{Four-Point Function}
\label{sec6}

The connected four point function is given by the contour ordered expectation value,
\bea
M^C_{abcd}(X,Y,Z,W) = \langle T_c \phi_a(X) \phi_b(Y) \phi_c(Z) \phi_d(W)\rangle \nonumber
\eea
The 1PI four-point function is obtained by truncating external legs and forms a 16 component 
tensor which we can write as the
outer product of four two component vectors,
\bea
M = {x \choose y} {u\choose v} {w\choose z} {s\choose t} \nonumber
\eea

The retarded 1PI four-point functions are given by
\begin{eqnarray}
M_{R1} &=& M_{1111} + M_{1112} + M_{1121} + M_{1211} + M_{1122} + M_{1212} +
M_{1221} + M_{1222} \nonumber\\
   &=& \frac{1}{2} (x-y)(u+v)(w+z)(s+t)\nonumber \\
M_{R2} &=& M_{1111} + M_{1112} + M_{1121} + M_{2111} + M_{1122} + M_{2112} +
M_{2121} + M_{2122}\nonumber \\
   &=& \frac{1}{2} (x+y)(u-v)(w+z)(s+t) \label{Ms}\\
M_{R3} &=& M_{1111} + M_{1112} + M_{2111} + M_{1211} + M_{2112} + M_{1212} +
M_{2211} + M_{2212} \nonumber\\
   &=& \frac{1}{2} (x+y)(u+v)(w-z)(s+t)\nonumber\\
M_{R4} &=& M_{1111} + M_{2111} + M_{1121} + M_{1211} + M_{2121} + M_{2211} +
M_{1221} + M_{2221} \nonumber\\
   &=& \frac{1}{2} (x+y)(u+v)(w+z)(s-t)\nonumber
\end{eqnarray}
where we have used the relation
\begin{equation}
\sum_{a,b,c,d=1}^{2} M_{abcd} = 0.
\end{equation}
The other combinations we will define as,
\begin{eqnarray}
M_A &=& \frac{1}{2}(x+y) (u+v) (w-z) (s-t)\nonumber \\
M_B &=& \frac{1}{2}(x-y) (u+v) (w-z) (s+t)\nonumber\\
M_C &=& \frac{1}{2}(x+y) (u-v) (w-z) (s+t) \nonumber\\
M_D &=& \frac{1}{2}(x+y) (u-v) (w+z) (s-t) \nonumber\\
M_E &=& \frac{1}{2}(x-y) (u-v) (w+z) (s+t) \nonumber\\
M_F &=& \frac{1}{2}(x-y) (u+v) (w+z) (s-t) \\
M_\alpha &=& \frac{1}{2}(x+y) (u-v) (w-z) (s-t)\nonumber\\
M_\beta &=& \frac{1}{2}(x-y) (u+v) (w-z) (s-t)\nonumber\\
M_\gamma &=& \frac{1}{2}(x-y) (u-v) (w+z) (s-t) \nonumber\\
M_\delta &=& \frac{1}{2}(x-y) (u-v) (w-z) (s+t) \nonumber\\
M_T &=& \frac{1}{2}(x-y) (u-v) (w-z) (s-t) \nonumber
\end{eqnarray}

We use the decomposition of the four point vertex:
\bea
8M &&= M_{R1}{1\choose -1}{1\choose 1}{1\choose 1}{1\choose 1} + 
M_{R2} {1\choose 1}{1\choose -1}{1\choose 1}{1\choose 1} + 
M_{R3} {1\choose 1}{1\choose 1}{1\choose -1}{1\choose 1} \nonumber \\
&&+ M_{R4} {1\choose 1}{1\choose 1}{1\choose 1}{1\choose -1} + 
M_A {1\choose 1}{1\choose 1}{1\choose -1}{1\choose -1} + 
M_B {1\choose -1}{1\choose 1}{1\choose -1}{1\choose 1} \nonumber\\
&&+ M_C {1\choose 1}{1\choose -1}{1\choose -1}{1\choose 1} + 
M_D {1\choose 1}{1\choose -1}{1\choose 1}{1\choose -1} + M_E {1\choose -1}{1\choose -1}{1\choose 1}{1\choose 1} \label{decompM}\\
&&+ M_F {1\choose -1}{1\choose 1}{1\choose 1}{1\choose -1} + 
M_{\alpha} {1\choose 1}{1\choose -1}{1\choose -1}{1\choose -1} + 
M_\beta {1\choose -1}{1\choose 1}{1\choose -1}{1\choose -1} \nonumber\\
&&+ M_\gamma {1\choose -1}{1\choose -1}{1\choose 1}{1\choose -1} + 
M_\delta {1\choose -1}{1\choose -1}{1\choose -1}{1\choose 1} + 
M_T {1\choose -1}{1\choose -1}{1\choose -1}{1\choose -1}\nonumber
\eea

\subsection{Rules for calculating Feynman Amplitudes}

The rules for handeling Keldysh indices are as follows: 
(for details see Ref.~\cite{PeterH}): 
\newline
1) Bare vertices carry a factor
\bea
\tau = {1 \choose -1} \label{tau}
\eea 
2) Internal indices are to be summed over. 
In terms of the column vectors ocurring in~(\ref{decompD}), (\ref{decompG})
and~(\ref{decompM}) this means that one adds the product of the 
upper components to the product of the lower components of all 
column vectors carrying the same internal index. The product of any number of vectors carrying 
the same internal index gives a scalar:
 \begin{equation}
   {x_1 \choose x_2} {x_3 \choose x_4} = x_1 x_3 + x_2 x_4 \, .
 \end{equation}
3) For external indices the product of any number of column vectors carrying the same index is 
defined to be another column vector whose upper (lower) component 
is given by the product of upper (lower) components of the original
vectors:
 \begin{equation}
   {x_1 \choose x_2} {x_3 \choose x_4} = { x_1 x_3 \choose x_2 x_4} \, .
 \end{equation}
\section{The Program}
The program was 
written entirely in $Mathematica$ 3.0.  $Mathematica$ was chosen because of
its powerful numeric algorithms and ability to perform operations on
sets \cite{wolfram}.  The program performs contractions on Keldysh indices.  It works 
for diagrams with three or four point interactions, with up to four external legs.
Any number of loops can be considered, and any number of the vertices can be corrected vertices.  Corrected vertices may be necessary when using an effective theory that involves a reorganized perturbation theory which is obtained by a resummation.  For example, in the hard thermal loop approximation, it is not sufficient to consider only bare vertices.  
All fields are treated as scalars and the coefficient is calculated by assuming that a bare three-point vertex carries a factor $-ig$, and a bare four-point vertex carries a factor $-i\lambda$.  Sign conventions for the 1PI functions are as shown in Fig. [1].  When non-scalar fields are involved, additional factors (such as traces over Dirac matrices for fermions, or contractions of projection operators for photons in a given gauge) must be calculated by hand.  Also note that each line carries a factor of $i$ which means that if gauge boson propagators are present, usual conventions require the insertion of an additional minus sign if there is an odd number of gauge boson propagators.  In addition, the sign for the gauge boson polarization tensor must be changed (see Fig. [1]).    
The user of the program must assign momenta to propagators and external legs so that the conservation of momentum is satisfied.  Keldysh indices must also be assigned.  When the program is executed, a number of input parameters
are requested.  We describe below the data entry process using the example shown in Fig. [2].  

\noindent {\bf 1)}. The number of external legs is entered (3).  For each
external leg, the momentum and index are recorded in the following form, 
\bea
\{P,a\}\,;~~~\{K,b\}\,;~~~\{PK,c\}\,.\nonumber
\eea
Note that the momentum is entered without signs or spaces.  The only purpose of these variables is to remind the user of the order he has chosen for the external legs.  For the example used in this section, the program will calculate $\Gamma(P,K,-P-K)$ since the order in which the external indices have been entered corresponds to this ordering of the momentum variables. 

\noindent {\bf 2)}. The number of internal indices (4) and a 
set containing these indices $\{d,e,f,g\}$ is entered.

\noindent {\bf 3)}. A set containing the loop variables is entered $\{R\}$.
This set contains the independent momenta, and is used when
the user wishes to have terms removed which are zero by contour
integration.  This point will be discussed further in section III-B.  

\noindent {\bf 4)}. The number of corrected three-point vertices is requested (1), and two further input boxes request the data 
pertaining to each corrected
vertex.  The first data group consists of a subscript, and the indices
for the vertex $\{1,e,a,f\}$.  The second group of data to be entered is the momenta corresponding to the previous indices $\{R,P,PR\}$.   As before, these momenta are simply used as a record keeping device.
To avoid clutter, the result is given without displaying explicitly the momentum dependence of the vertices.  To remind the user of the choices he has made, the momentum dependence for each vertex is printed out.  When there are more than one corrected three-point vertices, the subscripts distinguish them. 

\noindent {\bf 5)}. The number of corrected four-point vertices (1) and their parameters are
entered.  The format is the same as for the corrected three-point vertices: $\{1,d,b,c,g\}$, $\{R,K,PK,PR\}$.

\noindent {\bf 6)}. The number of indices that have a $\tau$ associated with them is entered (0).  If this number were non-zero, a further input box would request a set of these indices.  

\noindent {\bf 7)}. All of the data for each propagator is entered: the number of propagators (2), and each propagator's momentum, and initial and final index.  
In this case, the momenta are not merely used for book keeping purposes (as in the case of the external legs and corrected vertices).  The propagator momenta are used when eliminating terms that are zero by contour integration, and must be entered in a specific form. Each component of a propagator's momentum is entered seperately as a series of variables within a set bracket, and signs are included.  If there is only one term in the momentum, it must still 
be enclosed by set brackets:
\bea
\{\{R\},d,e\} \nonumber \\
\{\{P,R\},f,g\} \nonumber
\eea\ 

\noindent {\bf 8)}. The user is now asked if he would like the coefficient evaluated.  To calculate the coefficient, the number of bare three-point vertices (0) and bare
four-point vertices (0) must be entered.  For three- and four-point functions, the coefficient is calculated using the formula
\bea
{\rm coefficient} = \left(\frac{i}{2}\right)^p (-ig)^t(-i\lambda)^f \left(\frac{1}{4}\right)^{\Gamma}
\left(\frac{1}{8}\right)^{M} 2^{in-1}\nonumber
\eea
where $p$ is the number of propagators, $t$ is the number of bare three-point vertices, $f$ is the number of bare four-point vertices, $\Gamma$ is the number of corrected three-point vertices, $M$ is the number of corrected four-point vertices, and $in$ is the total number of indices.  For two-point functions, an extra factor of $-i$ is included (see Fig. [1]).

\noindent {\bf 9)}. At this point, the user is asked which combination of external indices he would like
evaluated.  For example, we will choose to evaluate the retarded combination $\Gamma_R(P,K,-P-K)$.

\noindent {\bf 10)}. At the user's discretion, the result of the calculation can 
be displayed with or without the terms removed which are zero by contour
integration. If the program is asked to remove these terms, it will remove terms in which, for any loop momentum, all poles of the propagators are on the same side of the real axis.  Since the complex integral can be evaluated by choosing a semi-circle in the upper or lower half plane, these terms normally give zero.  In some cases however (diagrams with tadpoles), there are terms with all poles are on the same side of the real axis which do not give zero (because of contributions from the semi-circle at infinity).  In addition, the program does not consider poles within the corrected vertices.  When corrected vertices are present, or if there are tadpole pieces to the diagram, the user must tell the program not to remove any of the terms that it thinks will be zero.  

\noindent {\bf 11)}. The user
can then choose to evaluate any other combination of external indices using 
the initial data.  For example, he could choose to calculate $\Gamma_{Fi}(P,K,-P-K)$.

The result is displayed in one of two ways.  If there are
less than four propagators, the result is simply shown as an unfactored
polynomial.  If there are four or more propagators, terms
involving the first two propagators entered are factored out of the
entire polynomial.  The $Mathematica$ operation $Simplify[\ldots ]$ is used on the
remaining terms.  
In addition, certain elements of the initial input
are echoed as output to help the user detect typing mistakes.  

For the example shown in Fig. [1] the result is;
\bea 
&&\Gamma_R(P,R,-P-R) \nonumber \\
&&~~~= -\frac{1}{2} \int dR \left\{ M_{R3}[r_r a_{p+r} \Gamma_{Fi} + r_r f_{p+r} \Gamma_R + f_r a_{p+r} \Gamma_{Ro}] + M_A a_r a_{p+r} \Gamma_{Ro} + M_B r_r r_{p+r} \Gamma_R \right\}\; \nonumber\\
&&M:= M(P+R,-P-K,K,-R)\,~~~\Gamma:=\Gamma(P,R,-P-R) \nonumber 
\eea
where we have used the notation $dR = d^4r/(2\pi)^4$, $r_r = D_R(R)$, $f_{p+r} = D_F(P+R)$ etc.
This example contains two propagators (each of which contains three terms), one corrected three-point function (which contains seven terms), and one corrected four-point function (which contains 15 terms).  
To do the calculation by hand, one would have to evaluate $3^2\cdot 7\cdot 15 = 945$ terms.  
In fact, only five terms are non-zero.  The program identifies these non-zero terms.  The use of a program of this type makes real time finite temperature calculations practical, and allows us to exploit the advantages of the real time formalism, one of which is the fact that one obtains physical green functions directly.

\section{The Ward Identity for QED in a Second Order HTL Effective Theory}

In a gauge theory, the Ward identities are a reflection of the gauge symmetry:  if the theory is invarient under gauge transformations, then the green functions of the theory obey the Ward identities.  It is well known that the QED Ward identities hold in a first order HTL effective theory.  This result is a consequence of the fact that the HTL theory respects gauge invarience.  In this section we will verify that the Ward identity for the three-point function and the electron self-energy is obeyed in a second order HTL effective theory.  We will show that the  Ward identity\cite{mhm} 
\bea
K_\mu \Gamma^\mu_R = -ie[\Sigma_A(P) - \Sigma_R(P+K)] \label{WI}
\eea
is satisfied by the diagrams shown in Fig. [3] and Fig. [4].  In these diagrams the solid dots are corrected vertices, which are obtained by adding the HTL vertex to the bare vertex, and the dotted lines are HTL propagators.  
We will work in real time and use the program described in section III to perform the summations over Keldysh indices.  This calculation would be prohibitively tedious using standard calculational techniques.  

  We use the Coulomb gauge.  The gauge boson propagator is given by,
\bea
&&D_{00} = -\frac{1}{k^2 - \Pi_{00}} \nonumber \\
&& D_{ij} = -\frac{(\delta_{ij} - k_i k_j / k^2)}{K^2 - \Pi_t}\,;~~~~\Pi_t = \frac{1}{2} \Pi_{ii} - \frac{k_0^2}{2k^2}\Pi_{00} \nonumber
\eea
where we use the notation $K_\mu =(k_0,\vec{k}),$ and $\Pi_{00}$ and $\Pi_{ii}$ are components of the HTL polarization tensor.  For simplicity, we will consider only longtitudinal modes.  To avoid the introduction of more notation, we will not explicitly distinguish these propagators from the bare propagators in section II. For example, we write
\bea
(D_{00}(K))_R = -\frac{1}{k^2 - \Pi_{00})_R}:= r_k \nonumber
\eea
and 
\bea
f_p = N^B_p[r_p-a_p]\,;~~~N^B_p = 1+2n(p_0)
\label{fb}
\eea
where $n(p_0)$ is the Bose-Einstein distribution function (\ref{BE}).
Fermion propagators are written with tildes.  
For example,
\bea
S^{-1}_R(P) = P\sla - \Sigma_R(P)\,;~~~~\tilde r_p = S_R(P)\nonumber
\eea
where $\Sigma(P)$ is the hard thermal loop fermion self-energy.  
The symmetric propagator is defined as \cite{mhm},
\bea 
\tilde f_p  =  N_F(P)P(\tilde r_p-\tilde a_p)\label{ff}
\eea
where $N_F(P)$ is constructed from the Fermi-Dirac distribution function,
\bea
N^F_p = 1-2n^f(p_0)\,;~~~~n^f(p_0) = \frac{1}{e^{\beta p_0} + 1}
\label{Nf}
\eea
To simplify notation, we suppress the Lorentz index $0$ on all vertices.  For example, for the three-point vertex we write $\Gamma_0:= \Gamma$.  For the four-point vertex we write $M_{\mu 0} := M_\mu$.  In addition, all Dirac indices are suppressed.

We will need to write four- and three-point functions with various momentum dependencies.  Suppressing all indices for the moment, we make the following definitions,
\bea
&&M^{(i)} := M(P+R,-P-K,K,-R) \nonumber \\
&&M^{(ii)} :=M(P,-P-K-R,K,R) \nonumber \\
&&\Gamma^{(A)} :=\Gamma(P,R,-P-R) \nonumber \\
&&\Gamma^{(B)} :=\Gamma(P+K+R,-R,-K-P) \label{defns} \\
&&\Gamma^{(C)} :=\Gamma(P+R,K,-P-K-R)  \nonumber 
\eea
We will also need vertices with the signs of the momenta reversed.  We use the following notation, 
\bea 
{\rm if}~~\Gamma^{(X)} = \Gamma(P1,P2,P3)~~{\rm then}~~\Gamma^{(X-)}=\Gamma(-P1,-P2,-P3)
\eea

The four- and three-point HTL corrected vertices are related through the Ward identities.  The identities that we will need are \cite{four-point},
\bea
&&K\cdot M_{R3}^{(i)} = e(\Gamma_{Ri}^{(A-)} - \Gamma_{Ri}^{(B)}) \nonumber \\
&&K\cdot M_{A}^{(i)} = e(\Gamma_{Fo}^{(A-)}) - \Gamma_{Fo}^{(B)}) \nonumber \\
&&K\cdot M_{B}^{(i)} = e\Gamma_{F}^{(A-)} \label{wd34} \\
&&K\cdot M_{R3}^{(ii)} = e(\Gamma_{Ro}^{(A)} - \Gamma_{Ro}^{(B-)}) \nonumber \\
&&K\cdot M_{A}^{(ii)} = e(\Gamma_{Fi}^{(A)} - \Gamma_{Fi}^{(B-)}) \nonumber \\
&&K\cdot M_{B}^{(ii)} = e\Gamma_{F}^{(A)}  \nonumber \\
&&K\cdot M_C^{ii} = -e\Gamma_F^{(B-)}\nonumber
\eea

Using the $Mathematica$ program described in section III, we calculate the contribution from the diagrams in Fig. [3A] and
Fig. [3B]:
\bea
\Gamma_{R}^{\mu(a)}(P,K, -P-K)=&&\frac{1}{2} \int dR\, \left[\right.(M_{R3}^{(i)})^\mu (r_r \tilde a_{p+r}\Gamma_{Fi}^A+r_r\tilde f_{p+r}\Gamma_{R}^A
+f_r \tilde a_{p+r} \Gamma_{Ro}^A) \nonumber\\
&& + (M_B^{(i)})^\mu  r_r \tilde r_{r+p} \Gamma^A_{R} + (M_A^{(i)})^\mu  a_r \tilde a_{r+p} \Gamma_{Ro}^A \left.\right]
\label{ga}\\
\Gamma_{R}^{\mu(b)}(P,K, -P-K)=&&\frac{1}{2}\int dR\,\left[\right. (M_{R3}^{(ii)})^\mu (a_r \tilde r_{p+k+r}\Gamma_{Fo}^B + a_r\tilde f_{p+k+r}\Gamma_{R}^B
+f_r \tilde r_{p+k+r} \Gamma_{Ri}^B) \nonumber \\
&& +(M_C^{(ii)})^\mu  a_r \tilde a_{r+k+p} \Gamma_{R}^B + (M_A^{(ii)})^\mu  r_r \tilde r_{r+k+p} \Gamma_{Ri}^B \left.\right]
\label{gb}
\eea

Using the Ward identity (\ref{wd34}) we obtain,
\bea
K^\mu(\Gamma_{R}^{\mu(a)} + \Gamma_{R}^{\mu(b)})=\frac{e}{2}\int dR\,(\alpha+\beta)
\label{alpha-p-beta}
\eea
where
\bea
\alpha&=&\Gamma_{Ri}^{(A-)}  (r_r \tilde a_{p+r} \Gamma_{Fi}^A + r_r\tilde f_{p+r}\Gamma_{R}^A
+f_r \tilde a_{p+r} \Gamma_{Ro}^A) + \Gamma_{F}^{(A-)}  r_r \tilde r_{r+p} \Gamma_{R}^A + \Gamma_{Fo}^{(A-)}  a_r \tilde a_{r+p} \Gamma_{Ro}^A \nonumber 
\\
&& -[ \Gamma_{Ro}^{B-} (a_r \tilde r_{p+k+r}\Gamma_{Fo}^B + a_r\tilde f_{p+k+r}\Gamma_{R}^B
+f_r \tilde r_{p+k+r} \Gamma_{Ri}^B) \label{alpha} \\
&&~~~~~~~~~~~~~~~~~~~~~~~~~~~~~~~~~~~~~~~~~~~~~~~~+ \Gamma_{F}^{(B-)}  a_r \tilde a_{r+p+k} \Gamma_{R}^B + \Gamma_{Fi}^{(B-)} r_r \tilde r_{r+p+k} \Gamma_{Ri}^B]  \nonumber\\
\beta &=& - \Gamma^B_{Ri} (r_r \tilde a_{p+r} \Gamma_{Fi}^A + r_r \tilde f_{p+r} \Gamma_R^A + f_r \tilde a_{p+r} \Gamma^A_{Ro}) - \Gamma^B_{Fo} a_r \tilde a_{p+r} \Gamma^A_{Ro} \nonumber \\
&&+\Gamma^A_{Ro}( a_r \tilde r_{p+k+r} \Gamma^B_{Fo} + a_r \tilde f_{p+k+r} \Gamma_R^B + f_r \tilde r_{p+k+r} \Gamma^B_{Ri}) + \Gamma^A_{Fi} r_k \tilde r_{p+k+r} \Gamma_{Ri}^B
\label{beta}
\eea

The $Mathematica$ program gives the retarded and advanced self-energies in Fig. [4] as,
\bea
\Sigma_R(P+K)&=&\frac{i}{2}\int dR\,\left[ \right.\Gamma_{Ro}^{(B-)}  (a_r \tilde r_{r+p+k}\Gamma_{Fo}^{B}+
a_r\tilde f_{r+p+k}\Gamma_{R}^{B}
+f_r \tilde r_{r+p+k} \Gamma_{Ri}^{B} ) \nonumber \\
&&~~~~+ \Gamma_{Fi}^{(B-)}  r_r \tilde r_{r+p+k} \Gamma_{Ri}^B + \Gamma_{F}^{(B-)}  a_r \tilde a_{r+p+k} \Gamma_{R}^B \left.\right]\nonumber
\\
\Sigma_A(P)&=&  \frac{i}{2} \int dR\,\left[\right. \Gamma_{Ri}^{(A-)}  ( r_r \tilde a_{p+r} \Gamma_{Fi}^A
+r_r\tilde f_{p+r}\Gamma_{R}^A
+f_r \tilde a_{p+r} \Gamma_{Ro}^A) \nonumber \\
&&~~~~+ \Gamma_{F}^{(A-)}  r_r \tilde r_{p+r} \Gamma_{R}^A + \Gamma_{F}^{(B-)} a_r \tilde a_{p+r} \Gamma_{R}^B \left.\right]
\label{piRA}
\eea
Comparing (\ref{alpha}) and (\ref{piRA}) we find that the contributions from the $\alpha$ term alone give, 
\bea
[K^\mu(\Gamma_{R}^{\mu(a)} + \Gamma_{R}^{\mu(b)})]_{\alpha~~{\rm only}} = -ie(\Sigma_A(P) - \Sigma_R(P+K))
\label{A}
\eea

Next we will show that the $\beta$ terms (\ref{beta}) cancel with the contributions from the diagram in Fig. [3C].
Using the $Mathematica$ program as before, we obtain the contribution from Fig. [3C].  The result is as follows:
\bea
\Gamma_{R}^{\mu(c)}&&(P,K, -P-K)= \frac{i}{2} \int dR\,\nonumber \\
\left[ \right.&& r_r \tilde a_{p+r} \tilde r_{p+k+r} (\Gamma^{(C)}_{R})_\mu \Gamma^{(A)}_{Fi} \Gamma^{(B)}_{Ri} + 
r_r \tilde f_{p+r} \tilde r_{p+k+r} (\Gamma^{(C)}_{R})_\mu \Gamma^{(A)}_{R } \Gamma^{(B)}_{Ri}\nonumber \\ 
&&+ 
a_r \tilde a_{p+r} \tilde r_{p+k+r} (\Gamma^{(C)}_{R})_\mu \Gamma^{(A)}_{Ro} \Gamma^{(B)}_{Fo} + 
a_r \tilde a_{p+r} \tilde f_{p+k+r} (\Gamma^{(C)}_{R})_\mu \Gamma^{(A)}_{Ro} \Gamma^{(B)}_{R}\label{gc} \\ 
&&+ 
f_r \tilde a_{p+r} \tilde r_{p+k+r} (\Gamma^{(C)}_{R})_\mu \Gamma^{(A)}_{Ro} \Gamma^{(B)}_{Ri} + 
r_r \tilde r_{p+r} \tilde r_{p+k+r} (\Gamma^{(C)}_{Fo})_\mu \Gamma^{(A)}_{R } \Gamma^{(B)}_{Ri}\nonumber \\ 
&&+ 
a_r \tilde a_{p+r} \tilde a_{p+k+r} (\Gamma^{(C)}_{Fi})_\mu \Gamma^{(A)}_{Ro} \Gamma^{(B)}_{R}\left.\right] 
\nonumber
\eea
Contracting with $K^\mu$ and using the Ward identities for the HTL corrected three point vertices:
\bea
K^\mu (\Gamma_R^{(C)})_\mu = -ie(\tilde r^{-1}_{p+k+r} - \tilde a^{-1}_{p+r}) \nonumber \\
K^\mu (\Gamma_{Fi}^{(C)})_\mu = ieN^F_{p+k+r} (\tilde a^{-1}_{p+k+r} - \tilde r^{-1}_{p+k+r}) \nonumber \\
K^\mu (\Gamma_R^{(C)})_\mu = ieN^F_{p+r}(\tilde r^{-1}_{p+r} - \tilde a^{-1}_{p+r}) \nonumber \\
\eea
we find that $K_\mu \Gamma^{\mu(c)}_R$ cancels exactly with (\ref{beta}).
As a result, we obtain from (\ref{A}),
\bea
\label{wd}
K^\mu(\Gamma^{(a)}_{R}+\Gamma^{(b)}_{R}+\Gamma^{(c)}_{R})_\mu=-ie(\Sigma_A(P)-\Sigma_R(P+K))
\eea
This is the Ward identity with full propagators and corrected vertices in QED
at finite temperature.

\section{Conclusion}

We have constructed a program with $Mathematica$ to evaluate Feynman amplitudes
in the Keldysh formalism of real time finite temperature field theory. This formalism
has recently gained increasing popularity because it avoids the need
for analytical continuations that plagues the imaginary formalism, and it allows for
a generalization to non-equilibrium situations. However, because of the extra degrees of 
freedom, calculations in the real time formalism can be extremely tedious,  especially for higher n-point functions. We have written a $Mathematica$ program that performs sums over Keldysh indices in the real time formalism.  The program calculates physical Feynman amplitudes for any diagram with up to
four external legs, with an arbitrary number of loops.  Generalization to diagrams
with more external legs is straightforward.  This program makes real time finite temperature calculations feasible for diagrams with  complicated structure. In  section III a diagram with 945 terms is calculated as an example.  The program performs the summations over Keldysh indices and produces a result in which only five terms are non-zero.

In order to demonstate the usefulness of this program, we have used it
to prove the finite temperature QED Ward identity in a second
order HTL effective theory. The relevant diagrams include corrected
three- and four-point vertices and full propagators. To calculate them by hand would be extremely time consuming. Using the program developed in this paper, we are able to calculate these diagrams and verify the QED Ward identity with a minimum of effort.

\section{Acknowledgements}

We are grateful to R. Kobes for useful discussions.

\newpage

\Large

\centerline{\bf Figure Captions}

\normalsize

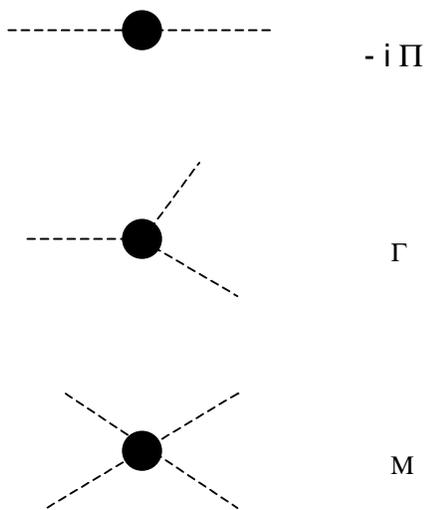
\noindent Fig. 1. Conventions for the definitions of the vertex funtions.

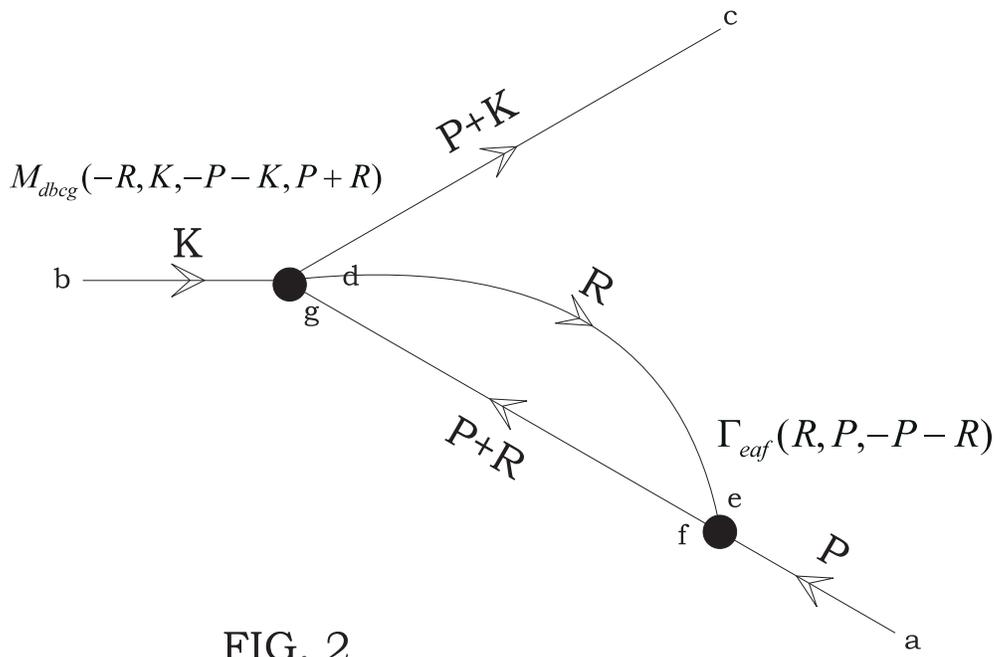
\noindent Fig. 2.  Example of a three-point function used to describe input parameters.

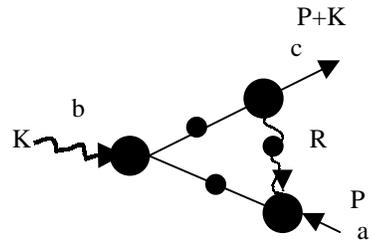
\noindent Fig. 3. Three-point functions that contribute to the Ward identity.

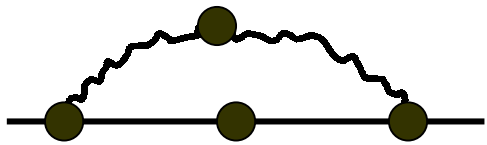
\noindent Fig. 4. Two-point function that contributes to the Ward identity.

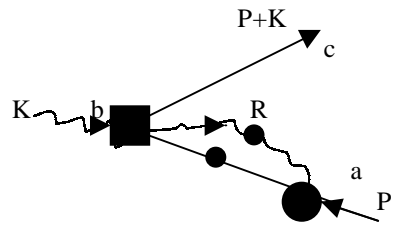
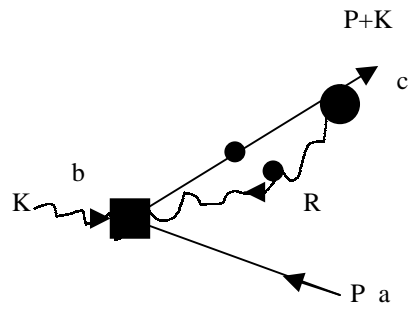
\end{document}